\documentclass[reprint,amsmath,amsfonts,amssymb,aps,prx,preprintnumbers,superscriptaddress]{revtex4-2}

\usepackage[utf8]{inputenc}
\usepackage[T1]{fontenc}
\usepackage{lmodern}
\usepackage{graphicx}
\usepackage{dcolumn}
\usepackage{bm}
\usepackage{textcomp}
\usepackage{ifpdf}
\usepackage{physics}
\usepackage[squaren,Gray]{SIunits}
\usepackage{color}
\definecolor{red}{rgb}{1,0,0}
\definecolor{blue}{rgb}{0,0,1}
\definecolor{darkred}{rgb}{0.6,0,0}
\definecolor{darkblue}{rgb}{0,0,0.6}
\definecolor{darkgreen}{rgb}{0,0.5,0}
\definecolor{grey}{rgb}{0.5,0.5,0.5}

\ifpdf
\usepackage{epstopdf}
\usepackage[pdftex,unicode,pdfstartview={FitH},pdfborder={0 0 0}]{hyperref}
\usepackage{hypcap}
\else
\usepackage[hypertex]{hyperref}
\fi
\hypersetup{
    bookmarksnumbered = true,
    colorlinks = true, linkcolor = darkblue,
    citecolor = darkblue, filecolor = darkblue,
    menucolor = darkblue, urlcolor = darkblue
}

\newcolumntype{R}{>{$\displaystyle}r<{$}}
\newcolumntype{C}{>{$\displaystyle}c<{$}}

\begin{document}

\title{Optimizing strong light-matter coupling of plasmonic lattices and monolayer semiconductors}

\author{Lukas Krelle}
\affiliation{Fakult\"at f\"ur Physik, Munich Quantum Center, and
Center for NanoScience (CeNS), Ludwig-Maximilians-Universit\"at
M\"unchen, Geschwister-Scholl-Platz 1, 80539 M\"unchen, Germany}
\affiliation{Institute for Condensed Matter Physics, Technische Universität Darmstadt, Hochschulstraße 6, 64289 Darmstadt, Germany }

\author{Lukas Husel}
\affiliation{Fakult\"at f\"ur Physik, Munich Quantum Center, and
Center for NanoScience (CeNS), Ludwig-Maximilians-Universit\"at
M\"unchen, Geschwister-Scholl-Platz 1, 80539 M\"unchen, Germany}
\affiliation{Department of Physics, University 
of Basel, 4056 Basel, Switzerland}

\author{Kenji Watanabe}
\affiliation{Research Center for Electronic and Optical Materials, National Institute for Materials Science, 1-1 Namiki, Tsukuba 305-0044, Japan}

\author{Takashi Taniguchi}
\affiliation{Research Center for Materials Nanoarchitectonics, National Institute for Materials Science,  1-1 Namiki, Tsukuba 305-0044, Japan}

\author{Ismail Bilgin}
\affiliation{Fakult\"at f\"ur Physik, Munich Quantum Center, and
Center for NanoScience (CeNS), Ludwig-Maximilians-Universit\"at
M\"unchen, Geschwister-Scholl-Platz 1, 80539 M\"unchen, Germany}
\affiliation{Faculty of Engineering, Department of Metallurgical and Materials Engineering, Karadeniz Technical University, Trabzon 61080, T\"urkiye}

\author{Alexander H\"ogele}
\email{alexander.hoegele@lmu.de}
\affiliation{Fakult\"at f\"ur Physik, Munich Quantum Center, and
  Center for NanoScience (CeNS), Ludwig-Maximilians-Universit\"at
  M\"unchen, Geschwister-Scholl-Platz 1, 80539 M\"unchen, Germany}
\affiliation{Munich Center for Quantum Science and Technology (MCQST),
  Schellingtra\ss{}e 4, 80799 M\"unchen, Germany}
  
\author{Farsane Tabataba-Vakili}
\email{f.tabataba@tu-braunschweig.de}
\affiliation{Fakult\"at f\"ur Physik, Munich Quantum Center, and
  Center for NanoScience (CeNS), Ludwig-Maximilians-Universit\"at
  M\"unchen, Geschwister-Scholl-Platz 1, 80539 M\"unchen, Germany}
\affiliation{Munich Center for Quantum Science and Technology (MCQST),
  Schellingtra\ss{}e 4, 80799 M\"unchen, Germany}
\affiliation{Institute of Condensed Matter Physics, Technische Universität Braunschweig, 38106 Braunschweig, Germany}
\affiliation{Laboratory for Emerging Nanometrology Braunschweig, 38106 Braunschweig, Germany}

\date{\today}

\begin{abstract}

Exciton-polaritons provide a versatile platform for the study of a wide range of phenomena, including polariton lasers, topological polaritons, and bosonic condensation. Transition metal dichalcogenide monolayers host excitons with large oscillator strength and binding energies constituting a robust matter constituent that forms polaritons from cryogenic to room temperature when embedded in optical microcavities. Plasmonic nanoparticles arranged in lattice geometries offer strong field-confinement and high quality factors. However, the high sensitivity of monolayer excitons to strain and dielectric disorder necessitates encapsulation in atomically flat hBN to ensure a high optical quality, rendering plasmonics more challenging. Here, we employ our recently developed fabrication method for embedding gold nanodisk arrays into van der Waals heterostructures and compare two samples with opposite layer order. We observe that strain and etching-induced surface contamination can reduce the exciton quality and thus the light-matter interaction strength significantly. Our fabrication approach reduces interfacial irregularities and enables homogeneous large-area polariton lattices for a wide range of applications, such as polarization-control or topological polaritonics.

\end{abstract}

\maketitle

\section{Introduction}

Exciton-polaritons enable a variety of phenomena, such as polariton lasing \cite{schneider2013electrically, harder2021coherent}, topological polaritons \cite{klembt2018exciton, liu2020generation, li2021experimental} and bosonic condensation \cite{kasprzak2006bose, anton2021bosonic, zhao2021ultralow}. Polaritons are quasi-particles that emerge when a matter excitation couples strongly to light inside a cavity. Among the variety of cavity systems employed for polaritonics, plasmonic surface lattice resonances (SLRs) have emerged as a versatile platform. These resonances appear when the localized surface plasmon resonance (LSPR) of a metal nanoparticle couples to the diffractive orders of the lattice, resulting in a collective excitation. Plasmonic SLRs combine the strong field confinement of LSPRs with high quality (Q) factors and have been used to realize photoluminescence (PL) enhancement \cite{lee2015fano}, lasing \cite{zhou2013lasing}, and polaritons \cite{liu2016strong, wang2019limits}. Owing to their large oscillator strengths and exciton binding energies, excitons in monolayer transition metal dichalcogenides (TMDs) are an ideal matter constituent for polaritons with demonstrations ranging from cryogenic \cite{liu2015strong, dufferwiel2015exciton, schneider2018two} to room temperature \cite{sun2017optical, zhang2018photonic, wang2019limits}. Due to their inherently two-dimensional (2D) character, excitons in TMDs exhibit a strong sensitivity to their environment, including strain \cite{khatibi2018impact} and dielectric disorder \cite{raja2019dielectric}. 

\begin{figure*}[t!]
\includegraphics[scale=1]{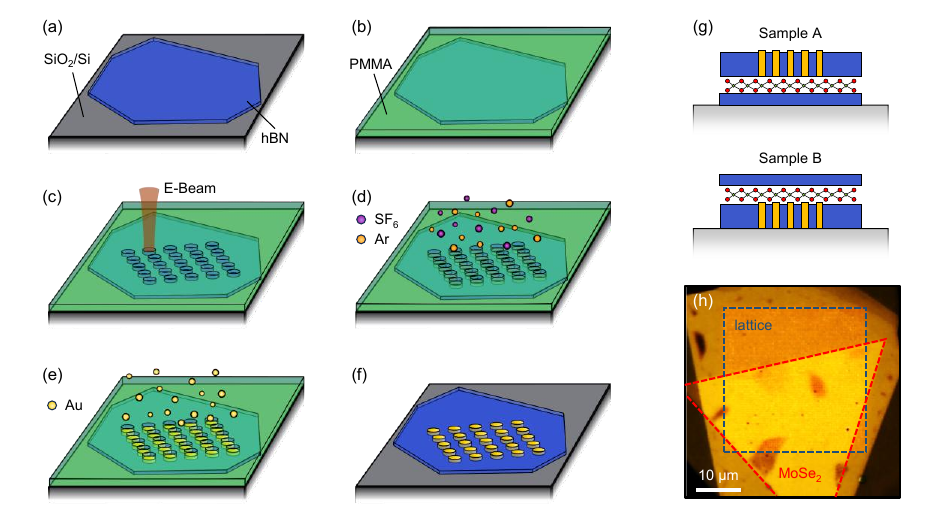}
\caption{\textbf{Fabrication of gold nanodisk arrays embedded in hBN.} Fabrication sequence: (a) mechanical exfoliation of an hBN flake on a SiO$_{2}$/Si substrate, (b) hBN  flake covered with a PMMA film, (c) electron-beam lithography, (d) ICP-RIE process of the lattice, (e) gold deposition and (f) removal of the PMMA film. (g) Schematics of the two sample geometries. (h) Optical  micrograph of sample A.} 
\label{fig:Fabrication}
\end{figure*}

Encapsulation of TMDs in high-quality hBN to form van der Waals heterostructures has become indispensable for samples with good optical quality, with the hBN serving as a clean and flat dielectric environment \cite{cadiz2017excitonic}. To couple emitters to a nanophotonic structure, such as plasmonic nanoparticles, a metasurface or a photonic crystal, the 2D material is commonly either placed on top of the nanostructure \cite{wang2019limits, sortino2019enhanced, vadia2023magneto, moilanen2025electrical} or the structure is  patterned directly on top of the material \cite{liu2016strong, liu2019observation, sun2021strong}. Additionally, recent advances were made in transferring nanostructures on top of an hBN carrier-layer onto an active medium \cite{dorey2025transferable}. However, in these schemes it is difficult to bring the emitter close to the near-field of the resonator without inducing strain and while ensuring encapsulation in hBN, since even a thin layer of hBN between emitter and resonator has a detrimental impact on the coupling strength. Recently, we reported on a novel fabrication method that addresses this issue by fully integrating a plasmonic cavity into an encapsulating hBN layer and thus into a van der Waals heterostructure providing direct proximity to the near-field without inducing strain \cite{tabataba2024metasurface}.

Here, we employ two fabrication schemes based on the method developed in Ref. \cite{tabataba2024metasurface}, comparing the influence of opposite layer order and using large-area MoSe$_2$ monolayers synthesized by chemical vapor deposition (CVD). In sample A, the hBN/plasmonic lattice was placed at the top of the heterostack, while it is at the bottom in sample B. We investigated exciton quality and homogeneity at cryogenic temperatures by differential reflectance (DR) and PL spectroscopy and found narrower linewidths and larger homogeneity of the exciton energy in sample A than in sample B. Using momentum-space DR and PL spectroscopy we further investigated the exciton-photon coupling and determined $25\%$ larger coupling strengths for polaritons in sample A than in sample B due to preserved oscillator strength and a strain-free environment of the exciton.

\begin{figure*}[t!]
\includegraphics[scale=1]{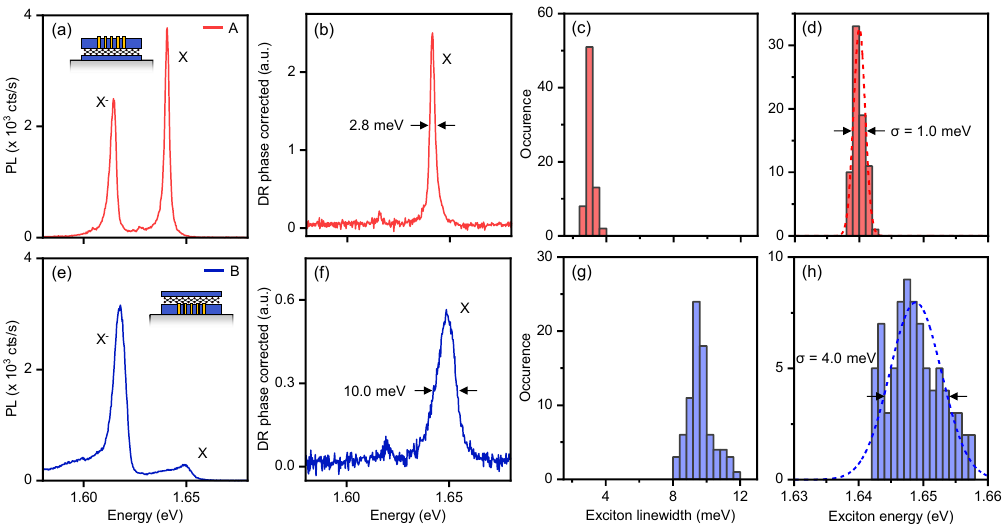}
\caption{\textbf{Monolayer characteristics.}  (a) Representative PL and (b) Kramers-Kronig phase-corrected DR spectra of sample A with the linewidth indicated by black arrows. In inset in (a) indicates the sample geometry. (c) Distribution of the exciton linewidth and (d) resonance energy in an isolated MoSe$_{2}$ region on sample A with dashed lines indicating a Gaussian distribution of the resonance energy and arrows marking the standard deviation. (e)-(h) Same as (a)-(d) but for sample B.} 
\label{fig:Exciton_Quality}
\end{figure*}

\section{Results and Discussion}

Plasmonic gold nanodisk arrays embedded in hBN were fabricated by employing electron-beam lithography (EBL) and inductively-coupled plasma reactive ion etching (ICP-RIE) in a stepwise fabrication process displayed in Figs.~\ref{fig:Fabrication}(a)-(f). HBN flakes were prepared via mechanical exfoliation from bulk crystals [Fig.~\ref{fig:Fabrication}(a)] and a square nanodisk array was defined via EBL using PMMA 950 K resist at an average dose of $500$\:$\mu$C/cm$^2$ [Figs.~\ref{fig:Fabrication}(b) and (c)]. Holes were etched through the hBN layers via ICP-RIE at an ICP power of $70$\:W and RF power of $6$\:W using $5$\:sccm of Ar and $10$\:sccm of SF$_{6}$ at a constant chamber pressure of $10$\:mTorr resulting in an etch rate of $0.6$\:nm/s [Fig.~\ref{fig:Fabrication}(d)]. After etching, a gold film matching the thickness of the hBN layer was deposited via electron-beam evaporation in ultra-high vacuum [Fig.~\ref{fig:Fabrication}(e)]. The lift-off was performed in acetone and isopropanol and the remaining residues were removed with an O$_{2}$ plasma [Fig.~\ref{fig:Fabrication}(f)]. The resulting top surface of the gold disks showed an irregular shape mimicking the etch profile, but the backside exhibited a flat and smooth surface \cite{tabataba2024metasurface}.

We fabricated two samples with opposite layer orders. Both samples consist of an hBN-embedded plasmonic lattice, a MoSe$_2$ monolayer and a second hBN encapsulation layer. In sample A, the hBN/lattice was placed on top of the stack, while it is the bottom layer in sample B, as shown in Fig.~\ref{fig:Fabrication}(g). Fig.~\ref{fig:Fabrication}(h) shows an optical micrograph of the final stack of sample A exhibiting only a small amount of lattice irregularities. In sample B, the elevated gold disks induce strain in the monolayer, whereas the bottom side of the disks provide a flat interface in sample A. Additionally, the top surface of the processed hBN layer was exposed to O$_{2}$ plasma, which induces defects \cite{comtet2019wide} and modifies the surface composition \cite{ma2019control}. The bottom surface of the hBN/lattice, on the other hand, was not exposed to any plasma treatment, and thus should provide a clean and flat interface for the MoSe$_2$ monolayer in sample A \cite{tabataba2024metasurface}.

We begin our comparison of the two sample schemes by investigating the matter constituents via confocal DR and PL spectroscopy at T = 4\:K in hBN-encapsulated monolayer regions in the two samples. Sample A displays emission from both neutral exciton and trion [see Fig.~\ref{fig:Exciton_Quality}(a)], with linewidths down to 2.4\:meV, close to the intrinsic limit of around 1\:meV \cite{moody2015intrinsic}, a clear sign of a pristine dielectric environment in spite of the additional fabrication steps. Sample B, on the other hand, exhibits a largely suppressed emission from the neutral exciton [Fig.~\ref{fig:Exciton_Quality}(e)] and is dominated by a strong trion emission, a sign of charge doping induced from O$_2$ plasma cleaning  \cite{kim2016effects, wu2019high}. Additionally, both emission lines exhibit asymmetries with broadened linewidths. Similarly to PL, sample A shows narrow absorption of the neutral exciton, whereas sample B shows a broadened absorption resonance with $\sim25\%$ lower oscillator strength   as determined by integrating over the peaks in the phase-corrected DR spectra of samples A and B in Figs.~\ref{fig:Exciton_Quality}(b) and (f), respectively. 

For a quantitative study, we analyzed the absorption of the neutral exciton over an area of $\sim 20~\mu \text{m}^2$ for both samples that we raster-scanned with a diffraction-limited optical spot. Figs.~\ref{fig:Exciton_Quality}(c) and (g) show the corresponding distributions of the exciton linewidths for samples A and B, respectively. Sample A displays an average narrow linewidth of 3.0 meV compared to the much broader linewidth of 9.6 meV in sample B. Similarly, the exciton energy is homogeneous with a standard deviation $\sigma_{X} \approx$ 1.0 meV around E$_{X} \approx$ 1.64 eV in sample A [Fig.~\ref{fig:Exciton_Quality}(d)] compared to $\sigma_{X} \approx$ 4.0 meV around E$_{X} \approx$ 1.649 eV in sample B [Fig.~\ref{fig:Exciton_Quality}(h)]. The narrow linewidths and spatial homogeneity emphasize the clean and homogeneous dielectric environment in sample A, while the data of sample B suggest increased interfacial contamination.

\begin{figure*}[t!]
\includegraphics[width = 17cm]{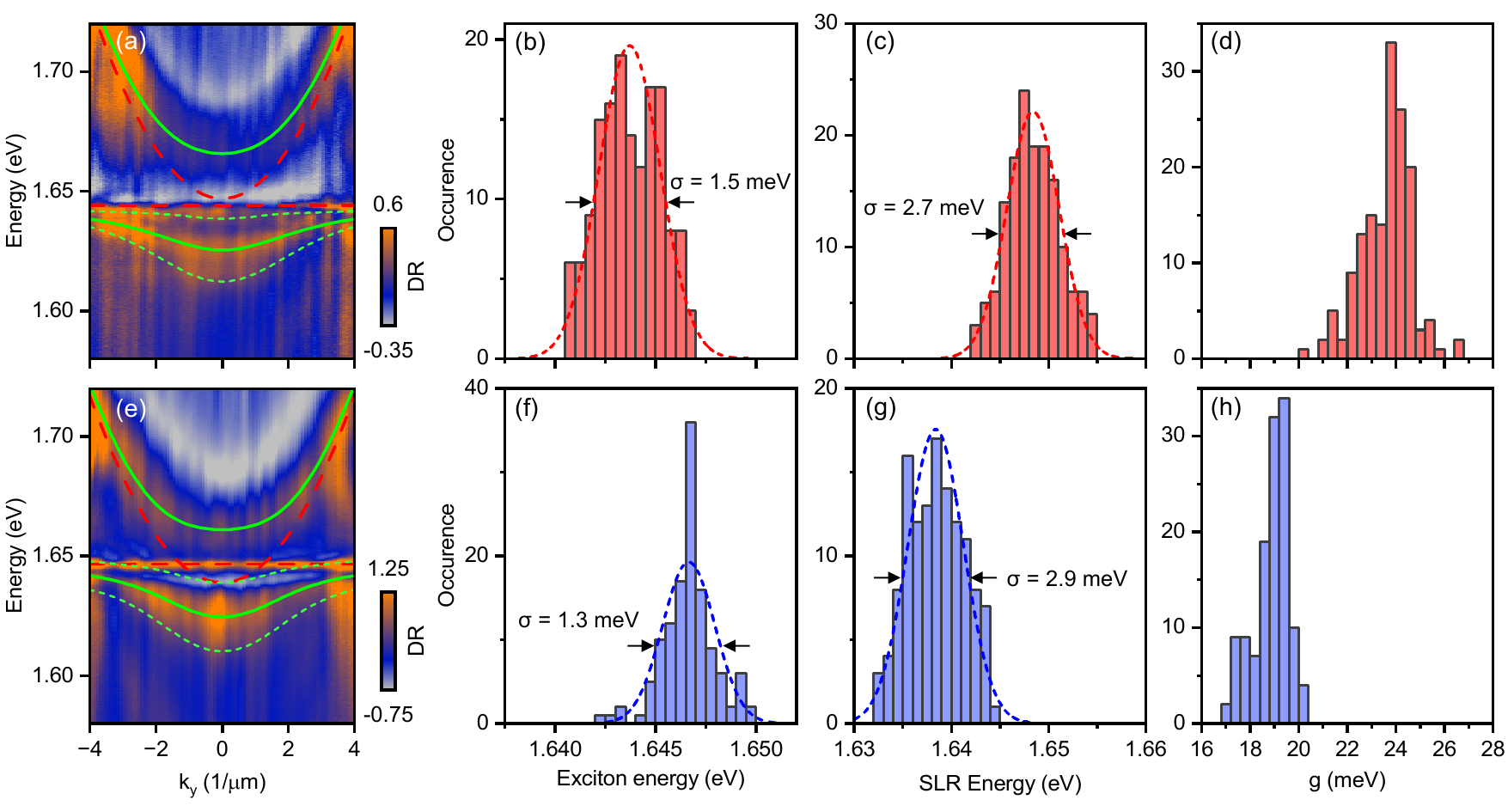}
\caption{\textbf{Exciton-plasmon coupling characteristics.} (a) Representative momentum-resolved polariton dispersion of sample A in DR with solid lines highlighting the fitted polariton branches and dashed lines indicating the exciton and SLR resonance energies. (b) Distribution of the resonance energy of the A1s exciton and (c) the SLR at $k = 0$ in sample A with dashed lines indicating a Gaussian distribution and the standard deviation marked by arrows. (d) Distribution of the exciton-plasmon coupling strength in sample A. (e)-(h) Same as (a)-(d) but for sample B.} 
\label{fig:Polariton_Quality}
\end{figure*}

For the investigation of light-matter coupling, we studied the momentum-resolved dispersions of the coupled exciton-plasmon systems, as shown in Figs.~\ref{fig:Polariton_Quality} (a) and (e) for the degenerate, hyperbolic lattice dispersions of samples A and B, respectively. The SLRs are well-described by a simple geometric model of the diffractive orders of the lattice \cite{guo2017geometry} (see Ref. \cite{tabataba2024metasurface} for a detailed discussion of this structure). We find that the layer order does not critically impact the linewidth of the SLR and extract linewidths and Q factors of 52.1\:meV and 32 for sample A, and 41.9\:meV and 40 for sample B.

In both samples, we observe avoided crossings as the cavity energy approaches the exciton energy, which is a hallmark of strong coupling and evidences the formation of exciton-polaritons, in addition to an uncoupled exciton fraction, as seen in Figs.~\ref{fig:Polariton_Quality}(a) and (e). This uncoupled exciton population is likely situated in areas between the gold disks and away from their respective maximum near-fields \cite{greten2024strong}.

We employed a coupled oscillator model \cite{savona1995quantum} to compare the exciton-photon coupling strengths in the two samples. The Hamiltonian reads
\begin{equation}
    H = 
    \begin{pmatrix}
        E_{\text{X}}-i\gamma_{\text{X}} & g \\
        g & E_{\text{SLR}}-i\gamma_{\text{SLR}}
    \end{pmatrix}
    ,
\end{equation}
where E$_{\text{X}}$ (E$_{\text{SLR}}$) and $\gamma_{\text{X}}$ ($\gamma_{\text{SLR}}$) are the exciton (SLR) energy and half-width at half-maximum and $g$ denotes the light-matter coupling strength. The energy eigenvalues for the upper (E$_{\text{UP}}$) and lower polariton (E$_{\text{LP}}$) branches obtained by diagonalizing $H$ are  
\begin{equation}\label{ham}
    \begin{split}
    E_{\text{UP,LP}} &= \frac{E_{\text{SLR}} + E_{\text{X}} - i(\gamma_{\text{X}}-\gamma_{\text{SLR}})}{2} \\
    &\pm \sqrt{g^{2} - \frac{1}{4}[E_{\text{X}} - E_{\text{SLR}} - i(\gamma_{\text{X}} - \gamma_{\text{SLR}})]^{2}}.
    \end{split}
\end{equation}

We fitted the dispersions in Figs.~\ref{fig:Polariton_Quality}(a) and (e) by tracking the DR maxima and found a higher coupling strength in sample A ($g = 23.5$\:meV) as opposed to sample B ($g = 19.5$\:meV), which we assign to the aforementioned difference in exciton oscillator strength [see Figs.~\ref{fig:Exciton_Quality}(b) and (f)]. 
The Rabi splitting, defined as $\Omega = \sqrt{4g^2-(\gamma_{\text{X}}-\gamma_{\text{SLR}})^2}$, is $40.1$ and $35.5$\:meV for samples A and B, respectively. In addition to the dispersions of the upper and lower polariton branches, we plot the linewidth of the lower polariton branch $\gamma_{LP}$ [see Fig.~\ref{fig:Polariton_Quality}(a) and (e)]. The linewidth is obtained from
\begin{equation}
    \gamma_{LP} = |X|^{2}\gamma_{X} + |C|^{2}\gamma_{SLR}
\end{equation}
with the Hopfield coefficients \cite{hopfield1958theory} given by
\begin{align}
    |X|^{2} &= \frac{1}{2} \Bigg[1 + \frac{E_{SLR} - E_{X}}{\sqrt{\Omega^{2} + (E_{SLR} - E_{X})^{2}}} \Bigg]\\
    |C|^{2} &= \frac{1}{2} \Bigg[1 - \frac{E_{SLR} - E_{X}}{\sqrt{\Omega^{2} + (E_{SLR} - E_{X})^{2}}} \Bigg].
\end{align}
We determined the exciton and SLR resonance energies, as well as the coupling strength across large areas of the two samples by fitting the coupled oscillator model to the data of every spatial position in 2D raster-scan momentum-space DR spectroscopy. The distribution of the exciton resonance energy at $k = 0$ of sample A  in Fig.~\ref{fig:Polariton_Quality}(b) shows a standard deviation $\sigma$ = 1.5 meV, comparable to the bare exciton region, highlighting the preserved clean environment for the exciton in spite of the presence of the lattice. Sample B displays a similar homogeneity with $\sigma$ = 1.3 meV in the hybrid region [see Fig.~\ref{fig:Polariton_Quality}(f)]. The SLR resonance energies in sample A and B, shown in Figs.~\ref{fig:Polariton_Quality}(c) and (g) respectively, exhibit comparable standard deviations for both samples. Hence, the layer order in the sample does not impact the homogeneity of the SLR resonance significantly, similar to the linewidth. Figs.~\ref{fig:Polariton_Quality}(d) and (h) display the distributions of the obtained coupling strengths $g$ for samples A and B, respectively. We obtain average $g$ values and standard deviations of $23.6\pm1$ and $18.9\pm0.7$\:meV, as well as maximum $g$ values of $26.4$ and $20.2$\:meV for samples A and B, respectively, demonstrating elevated coupling strengths in sample A over large areas. This is consistent with the results obtained for the bare exciton with larger oscillator strength in this sample.

\section{\label{sec:Conclusion}Conclusion}

In summary, we studied exciton-polaritons in samples with plasmonic SLRs coupled to MoSe$_2$ monolayers. Using a fabrication method that allows to embed plasmonic lattices into thin hBN layers and thus integrate them into van der Waals heterostructures \cite{tabataba2024metasurface}, we compared two samples with opposite layer order. We found that placing the hBN-embedded plasmonic lattice on top of the stack yielded high quality, narrow excitonic absorption and PL, while fabrication-induced surface irregularities and doping resulted in broader exciton resonances with reduced oscillator strength when the lattice was placed at the bottom of the stack. Momentum-resolved DR dispersions at $4$\:K exhibited avoided crossings, demonstrating strong exciton-photon coupling in both geometries. Fitting a coupled oscillator model across large areas revealed a $25\%$ increased coupling strength for the sample, in which the hBN-embedded plasmonic lattice was placed on top of the TMD monolayer, resulting in a clean and strain-free dielectric environment and preserved oscillator strength. Beyond the comparison of layer order presented here, our fabrication method establishes a versatile platform for exploring integration of sensitive van der Waals materials into nanophotonic device architectures with potential applications in polariton circuitry and topological photonics \cite{wang2016existence, honari2019topological, liu2020generation}.

\section{Methods}

\subsection{Synthesis of MoSe$_2$ monolayers} 

We synthesized large-area high-quality MoSe$_2$ monolayers using a CVD system (Carbolite Gero) with a three-zone furnace and a 1-inch quartz tube. The vapor phase chalcogenization method was used with MoO$_2$ and Se powders (99.99\%, Sigma-Aldrich) as precursors \cite{bilgin2018resonant}. The target substrates, Si chips with $285$\:nm SiO$_2$, were suspended face-down over an aluminum crucible with MoO$_2$ powder placed at the center of the first heating zone. $10$\:cm upstream from the center, another crucible with Se powder was placed. The tube was evacuated to $10$\:mTorr several times to remove air and moisture and $300$\:sccm of ultra-high purity Ar was flowed into the chamber to act as a carrier gas at ambient pressure in an inert atmosphere. For the growth, the furnace temperature was increased at a rate of $10$°C/min to $750$°C and maintained for $15$\:min. When the temperature reached $750$°C, $0.75$\:sccm of H$_2$ was introduced as a reductant gas. After the growth, the furnace was cooled down to room temperature.

\subsection{Sample design and assembly}

We employed the finite-difference time-domain method (software Lumerical) to design the gold nanodisk arrays. We simulated gold nanodisks with diameter of 70 nm organized on a square lattice with a lattice constant of $500$\:nm. The heights of the disks were designed between $30$ and $50$\:nm matching the thickness of the hBN layer, in which they were incorporated. We then selected the thickness of the second hBN layer such that the simulated resonance energy matched the energy of the neutral 1s exciton in the MoSe$_{2}$ monolayer.\\
For the sample assembly, we used a polymer assisted dry stamping method based on poly-bisphenol A-carbonate (PC, Sigma Aldrich) and polycaprolacton (PCL, Sigma Aldrich). For the pick-up of the lattices, PCL stamps yielded a higher success rate and reproducibility compared to PC. The transfer with PCL was performed by heating to 57$^{\circ}$C after contact, below the melting point (60$^{\circ}$C) \cite{son2020strongly}, followed by a pick-up at 30$^{\circ}$C. For sample B, we released the hBN/lattice flake onto the target substrate at 75$^{\circ}$C and cleaned it in THF, acetone, and isopopanol. The top hBN and CVD-grown MoSe$_2$ flakes were picked up successively with a PC stamp at $95$°C and the stack was released onto the hBN/lattice flake at $195$°C. For sample A, the hBN/lattice flake was released onto a CVD-grown MoSe$_{2}$ monolayer at 75$^{\circ}$C and cleaned in THF, acetone, and isopropanol. Then a PC stamp was used to pick up the stack at $95$°C, followed by release onto a pre-transferred (with PC) bottom hBN flake at $195$°C. PC was removed in chloroform, acetone, and isopropanol. Both samples were annealed in ultrahigh vacuum at $200$°C for $15$\:h to remove residues and trapped air from the interfaces.

\subsection{Optical spectroscopy}
We performed cryogenic DR and PL spectroscopy in backscattering geometry in a closed-cycle cryostat (attocube systems, attoDRY800) operating at a temperature of $4$\:K. Piezo-units (attocube systems, ANPx101, ANPz101, and ANSxy100) were used to position the sample with respect to a low temperature apochromatic objective (LT-APO/633-RAMAN/0.81). A lab-built Fourier imaging setup in 4f and telescope configuration employing four achromatic doublet lenses (Edmund Optics, VIS-NIR with focal lengths of $750$, $750$, $400$ and $150$\:mm) was used to measure angle-resolved DR and PL spectra. A spectrometer (Teledyne Princeton Instruments, IsoPlane SCT320) with a $300$\:grooves/mm grating was used to disperse the signal, which was detected by a Peltier-cooled charge-coupled device (CCD) (Teledyne Princeton Instruments, PIXIS 1024). We converted camera pixel to wavevector in 1/$\mu$m using the relation $k_{\parallel}=\text{sin}(\Theta)k_{0}$ with the maximum angle $\Theta$ given by the numerical aperture of the objective, the illuminated area on the CCD, and the light wavevector $k_{0}=2\pi \backslash \lambda$. For DR, we used a pulsed supercontinuum laser (NKT Photonics, SuperK with Varia filter). DR was calculated using the reflectance R from the lattice or hybrid region and the background R$_{0}$ from a region containing only hBN and was defined as $DR = (R-R_0)/R_0$. Exciton resonance linewidths and oscillator strengths were extracted using Kramers-Kronig relation and Lorentzian fits. PL was excited with an external cavity diode laser (New Focus, Velocity) emitting at $635$\:nm with excitation powers ranging from 7 $\mu$W to 30$\mu$W.

\vspace{8pt}
\noindent \textbf{Acknowledgements:} This research was funded by the European Research Council (ERC) under the Grant Agreement No.~772195 and the Deutsche Forschungsgemeinschaft (DFG, German Research Foundation) within Germany's Excellence Strategy under grant No.~EXC-2111-390814868. K.W. and T.T. acknowledge support from the CREST (JPMJCR24A5), JST and World Premier International Research Center Initiative (WPI), MEXT, Japan. I.\,B. acknowledges funding from the Alexander von Humboldt Foundation. F.\,T.-V. acknowledges funding from the Munich Center for Quantum Science and Technology (MCQST) and the European Union's Framework Programme for Research and Innovation Horizon Europe under the Marie Sk{\l}odowska-Curie Actions grant agreement No.~101058981.

\vspace{8pt}


\bibliography{bib_file}

@article{savona1995quantum,
  title={Quantum well excitons in semiconductor microcavities: Unified treatment of weak and strong coupling regimes},
  author={Savona, Vincenzo and Andreani, LC and Schwendimann, P and Quattropani, A},
  journal={Solid State Commun.},
  volume={93},
  number={9},
  pages={733--739},
  year={1995},
  publisher={Elsevier}
}

@article{tabataba2024metasurface,
  title={Metasurface of strongly coupled excitons and nanoplasmonic arrays},
  author={Tabataba-Vakili, Farsane and Krelle, Lukas and Husel, Lukas and Nguyen, Huy PG and Li, Zhijie and Bilgin, Ismail and Watanabe, Kenji and Taniguchi, Takashi and H{\"o}gele, Alexander},
  journal={Nano Lett.},
  volume={24},
  number={33},
  pages={10090--10097},
  year={2024},
  publisher={ACS Publications}
}

@article{honari2019topological,
  title={Topological plasmonic edge states in a planar array of metallic nanoparticles},
  author={Honari-Latifpour, Mostafa and Yousefi, Leila},
  journal={Nanophotonics},
  volume={8},
  number={5},
  pages={799--806},
  year={2019},
  publisher={De Gruyter}
}

@article{wang2016existence,
  title={The existence of topological edge states in honeycomb plasmonic lattices},
  author={Wang, Li and Zhang, Ruo-Yang and Xiao, Meng and Han, Dezhuan and Chan, Che Ting and Wen, Weijia},
  journal={N. J. Phys.},
  volume={18},
  number={10},
  pages={103029},
  year={2016},
  publisher={IOP Publishing}
}

@article{dufferwiel2015exciton,
  title={Exciton--polaritons in van der {W}aals heterostructures embedded in tunable microcavities},
  author={Dufferwiel, S and Schwarz, S and Withers, F and Trichet, AAP and Li, F and Sich, M and Del Pozo-Zamudio, O and Clark, C and Nalitov, A and Solnyshkov, DD and G. Malpuech and K. S. Novoselov and J. M. Smith and M. S. Skolnick and D. N. Krizhanovskii and A. I. Tartakovskii},
  journal={Nat. Commun.},
  volume={6},
  number={1},
  pages={8579},
  year={2015},
  publisher={Nature Publishing Group UK London}
}

@article{sun2021strong,
  title={Strong plasmon--exciton coupling in transition metal dichalcogenides and plasmonic nanostructures},
  author={Sun, Jiawei and Li, Yang and Hu, Huatian and Chen, Wen and Zheng, Di and Zhang, Shunping and Xu, Hongxing},
  journal={Nanoscale},
  volume={13},
  number={8},
  pages={4408--4419},
  year={2021},
  publisher={Royal Society of Chemistry}
}

@article{liu2019observation,
  title={Observation and active control of a collective polariton mode and polaritonic band gap in few-layer {WS}$_2$ strongly coupled with plasmonic lattices},
  author={Liu, Wenjing and Wang, Yuhui and Zheng, Biyuan and Hwang, Minsoo and Ji, Zhurun and Liu, Gerui and Li, Ziwei and Sorger, Volker J and Pan, Anlian and Agarwal, Ritesh},
  journal={Nano Lett.},
  volume={20},
  number={1},
  pages={790--798},
  year={2019},
  publisher={ACS Publications}
}

@article{vadia2023magneto,
  title={Magneto-Optical Chirality in a Coherently Coupled Exciton--Plasmon System},
  author={Vadia, Samarth and Scherzer, Johannes and Watanabe, Kenji and Taniguchi, Takashi and H{\"o}gele, Alexander},
  journal={Nano Lett.},
  volume={23},
  number={2},
  pages={614--618},
  year={2023},
  publisher={ACS Publications}
}

@article{sortino2019enhanced,
  title={Enhanced light-matter interaction in an atomically thin semiconductor coupled with dielectric nano-antennas},
  author={Sortino, Luca and Zotev, PG and Mignuzzi, S and Cambiasso, J and Schmidt, D and Genco, A and A{\ss}mann, M and Bayer, M and Maier, SA and Sapienza, R and A. I. Tartakovskii},
  journal={Nat. Commun.},
  volume={10},
  number={1},
  pages={5119},
  year={2019},
  publisher={Nature Publishing Group UK London}
}

@article{anton2021bosonic,
  title={Bosonic condensation of exciton--polaritons in an atomically thin crystal},
  author={Anton-Solanas, Carlos and Waldherr, Maximilian and Klaas, Martin and Suchomel, Holger and Harder, Tristan H and Cai, Hui and Sedov, Evgeny and Klembt, Sebastian and Kavokin, Alexey V and Tongay, Sefaattin and  Kenji Watanabe and Takashi Taniguchi and Sven H{\"o}fling and Christian Schneider},
  journal={Nat. Mater.},
  volume={20},
  number={9},
  pages={1233--1239},
  year={2021},
  publisher={Nature Publishing Group UK London}
}

@article{schneider2013electrically,
  title={An electrically pumped polariton laser},
  author={Schneider, Christian and Rahimi-Iman, Arash and Kim, Na Young and Fischer, Julian and Savenko, Ivan G and Amthor, Matthias and Lermer, Matthias and Wolf, Adriana and Worschech, Lukas and Kulakovskii, Vladimir D and Ivan A. Shelykh and Martin Kamp and Stephan Reitzenstein and Alfred Forchel and Yoshihisa Yamamoto and Sven H{\"o}fling},
  journal={Nature},
  volume={497},
  number={7449},
  pages={348--352},
  year={2013},
  publisher={Nature Publishing Group UK London}
}

@article{wang2019limits,
  title={Limits to strong coupling of excitons in multilayer {WS}$_2$ with collective plasmonic resonances},
  author={Wang, Shaojun and Le-Van, Quynh and Vaianella, Fabio and Maes, Bjorn and Eizagirre Barker, Simone and Godiksen, Rasmus H and Curto, Alberto G and Gomez Rivas, Jaime},
  journal={ACS Photon.},
  volume={6},
  number={2},
  pages={286--293},
  year={2019},
  publisher={ACS Publications}
}

@article{guo2017geometry,
  title={Geometry dependence of surface lattice resonances in plasmonic nanoparticle arrays},
  author={Guo, Rui and Hakala, Tommi K and T{\"o}rm{\"a}, P{\"a}ivi},
  journal={Phys. Rev. B},
  volume={95},
  number={15},
  pages={155423},
  year={2017},
  publisher={APS}
}

@article{bilgin2018resonant,
  title={Resonant Raman and exciton coupling in high-quality single crystals of atomically thin molybdenum diselenide grown by vapor-phase chalcogenization},
  author={Bilgin, Ismail and Raeliarijaona, Aldo S and Lucking, Michael C and Hodge, Sebastian Cooper and Mohite, Aditya D and de Luna Bugallo, Andres and Terrones, Humberto and Kar, Swastik},
  journal={ACS Nano},
  volume={12},
  number={1},
  pages={740--750},
  year={2018},
  publisher={ACS Publications}
}

@article{cadiz2017excitonic,
  title={Excitonic Linewidth Approaching the Homogeneous Limit in {M}o{S}$_{2}$-based van der {W}aals Heterostructures},
  author={Cadiz, Fabian and Courtade, Emmanuel and Robert, C{\'e}dric and Wang, Gang and Shen, Yuxia and Cai, Hui and Taniguchi, Takashi and Watanabe, Kenji and Carrere, Helene and Lagarde, Delphine and M. Manca and T. Amand and P. Renucci and S. Tongay and X. Marie and B. Urbaszek},
  journal={Phys. Rev. X},
  volume={7},
  number={2},
  pages={021026},
  year={2017},
  publisher={APS}
}

@article{klembt2018exciton,
  title={Exciton-polariton topological insulator},
  author={Klembt, Sebastian and Harder, TH and Egorov, OA and Winkler, K and Ge, R and Bandres, MA and Emmerling, M and Worschech, L and Liew, TCH and Segev, M and Schneider, C and H{\"o}fling, S},
  journal={Nature},
  volume={562},
  number={7728},
  pages={552--556},
  year={2018},
  publisher={Nature Publishing Group UK London}
}

@article{li2021experimental,
  title={Experimental observation of topological {Z}$_2$ exciton-polaritons in transition metal dichalcogenide monolayers},
  author={Li, Mengyao and Sinev, Ivan and Benimetskiy, Fedor and Ivanova, Tatyana and Khestanova, Ekaterina and Kiriushechkina, Svetlana and Vakulenko, Anton and Guddala, Sriram and Skolnick, Maurice and Menon, Vinod M and Krizhanovskii, Dmitry and Al{\`u}, Andrea and Samusev, Anton and Khanikaev, Alexander B},
  journal={Nat. Commun.},
  volume={12},
  number={1},
  pages={4425},
  year={2021},
  publisher={Nature Publishing Group UK London}
}

@article{liu2020generation,
  title={Generation of helical topological exciton-polaritons},
  author={Liu, Wenjing and Ji, Zhurun and Wang, Yuhui and Modi, Gaurav and Hwang, Minsoo and Zheng, Biyuan and Sorger, Volker J and Pan, Anlian and Agarwal, Ritesh},
  journal={Science},
  volume={370},
  number={6516},
  pages={600--604},
  year={2020},
  publisher={American Association for the Advancement of Science}
}

@article{kasprzak2006bose,
  title={Bose--{E}instein condensation of exciton polaritons},
  author={Kasprzak, Jacek and Richard, Murielle and Kundermann, S and Baas, A and Jeambrun, P and Keeling, Jonathan Mark James and Marchetti, FM and Szyma{\'n}ska, MH and Andr{\'e}, R and Staehli, JL and Savona, V and Littlewood, PB and Deveaud, B and Dang, Le Si},
  journal={Nature},
  volume={443},
  number={7110},
  pages={409--414},
  year={2006},
  publisher={Nature Publishing Group UK London}
}

@article{zhao2021ultralow,
  title={Ultralow Threshold Polariton Condensate in a Monolayer Semiconductor Microcavity at Room Temperature},
  author={Zhao, Jiaxin and Su, Rui and Fieramosca, Antonio and Zhao, Weijie and Du, Wei and Liu, Xue and Diederichs, Carole and Sanvitto, Daniele and Liew, Timothy CH and Xiong, Qihua},
  journal={Nano Lett.},
  volume={21},
  number={7},
  pages={3331--3339},
  year={2021},
  publisher={ACS Publications}
}

@article{harder2021coherent,
  title={Coherent Topological Polariton Laser},
  author={Harder, Tristan H and Sun, Meng and Egorov, Oleg A and Vakulchyk, Ihor and Beierlein, Johannes and Gagel, Philipp and Emmerling, Monika and Schneider, Christian and Peschel, Ulf and Savenko, Ivan G and Klembt, Sebastian and H{\"o}fling, Sven},
  journal={ACS Photon.},
  volume={8},
  number={5},
  pages={1377--1384},
  year={2021},
  publisher={ACS Publications}
}

@article{raja2019dielectric,
  title={Dielectric disorder in two-dimensional materials},
  author={Raja, Archana and Waldecker, Lutz and Zipfel, Jonas and Cho, Yeongsu and Brem, Samuel and Ziegler, Jonas D and Kulig, Marvin and Taniguchi, Takashi and Watanabe, Kenji and Malic, Ermin and Heinz, Tony F and Berkelbach, Timothy C and Chernikov, Alexey},
  journal={Nat. Nanotechnol.},
  volume={14},
  number={9},
  pages={832--837},
  year={2019},
  publisher={Nature Publishing Group UK London}
}

@article{moody2015intrinsic,
  title={Intrinsic homogeneous linewidth and broadening mechanisms of excitons in monolayer transition metal dichalcogenides},
  author={Moody, Galan and Kavir Dass, Chandriker and Hao, Kai and Chen, Chang-Hsiao and Li, Lain-Jong and Singh, Akshay and Tran, Kha and Clark, Genevieve and Xu, Xiaodong and Bergh{\"a}user, Gunnar and Malic, Ermin and Knorr, Andreas and Li, Xiaoqin},
  journal={Nat. Commun.},
  volume={6},
  number={1},
  pages={8315},
  year={2015},
  publisher={Nature Publishing Group UK London}
}

@article{zhou2013lasing,
  title={Lasing action in strongly coupled plasmonic nanocavity arrays},
  author={Zhou, Wei and Dridi, Montacer and Suh, Jae Yong and Kim, Chul Hoon and Co, Dick T and Wasielewski, Michael R and Schatz, George C and Odom, Teri W},
  journal={Nat. Nanotechnol.},
  volume={8},
  number={7},
  pages={506--511},
  year={2013},
  publisher={Nature Publishing Group UK London}
}

@article{liu2016strong,
  title={Strong exciton--plasmon coupling in {M}o{S}$_{2}$ coupled with plasmonic lattice},
  author={Liu, Wenjing and Lee, Bumsu and Naylor, Carl H and Ee, Ho-Seok and Park, Joohee and Johnson, AT Charlie and Agarwal, Ritesh},
  journal={Nano Lett.},
  volume={16},
  number={2},
  pages={1262--1269},
  year={2016},
  publisher={ACS Publications}
}

@article{lee2015fano,
  title={Fano resonance and spectrally modified photoluminescence enhancement in monolayer {M}o{S}$_{2}$ integrated with plasmonic nanoantenna array},
  author={Lee, Bumsu and Park, Joohee and Han, Gang Hee and Ee, Ho-Seok and Naylor, Carl H and Liu, Wenjing and Johnson, AT Charlie and Agarwal, Ritesh},
  journal={Nano Lett.},
  volume={15},
  number={5},
  pages={3646--3653},
  year={2015},
  publisher={ACS Publications}
}

@article{khatibi2018impact,
  title={Impact of strain on the excitonic linewidth in transition metal dichalcogenides},
  author={Khatibi, Zahra and Feierabend, Maja and Selig, Malte and Brem, Samuel and Linder{\"a}lv, Christopher and Erhart, Paul and Malic, Ermin},
  journal={2D Mater.},
  volume={6},
  number={1},
  pages={015015},
  year={2018},
  publisher={IOP Publishing}
}

@article{schneider2018two,
  title={Two-dimensional semiconductors in the regime of strong light-matter coupling},
  author={Schneider, Christian and Glazov, Mikhail M and Korn, Tobias and H{\"o}fling, Sven and Urbaszek, Bernhard},
  journal={Nat. Commun.},
  volume={9},
  number={1},
  pages={2695},
  year={2018},
  publisher={Nature Publishing Group UK London}
}

@article{liu2015strong,
  title={Strong light--matter coupling in two-dimensional atomic crystals},
  author={Liu, Xiaoze and Galfsky, Tal and Sun, Zheng and Xia, Fengnian and Lin, Erh-chen and Lee, Yi-Hsien and K{\'e}na-Cohen, St{\'e}phane and Menon, Vinod M},
  journal={Nat. Photon.},
  volume={9},
  number={1},
  pages={30--34},
  year={2015},
  publisher={Nature Publishing Group UK London}
}

@article{sun2017optical,
  title={Optical control of room-temperature valley polaritons},
  author={Sun, Zheng and Gu, Jie and Ghazaryan, Areg and Shotan, Zav and Considine, Christopher R and Dollar, Michael and Chakraborty, Biswanath and Liu, Xiaoze and Ghaemi, Pouyan and K{\'e}na-Cohen, St{\'e}phane and Menon, Vinod M},
  journal={Nat. Photon.},
  volume={11},
  number={8},
  pages={491--496},
  year={2017},
  publisher={Nature Publishing Group UK London}
}

@article{zhang2018photonic,
  title={Photonic-crystal exciton-polaritons in monolayer semiconductors},
  author={Zhang, Long and Gogna, Rahul and Burg, Will and Tutuc, Emanuel and Deng, Hui},
  journal={Nat. Commun.},
  volume={9},
  number={1},
  pages={713},
  year={2018},
  publisher={Nature Publishing Group UK London}
}

@article{son2020strongly,
  title={Strongly adhesive dry transfer technique for van der {W}aals heterostructure},
  author={Son, Suhan and Shin, Young Jae and Zhang, Kaixuan and Shin, Jeacheol and Lee, Sungmin and Idzuchi, Hiroshi and Coak, Matthew J and Kim, Hwangsun and Kim, Jangwon and Kim, Jae Hoon and Kim, Miyoung and Kim, Dohun and Kim, Philip and Park, Je--Geun},
  journal={2D Mater.},
  volume={7},
  number={4},
  pages={041005},
  year={2020},
  publisher={IOP Publishing}
}

@article{ma2019control,
  title={Control of hexagonal boron nitride dielectric thickness by single layer etching},
  author={Ma, Zichao and Prawoto, Clarissa and Ahmed, Zubair and Xiao, Ying and Zhang, Lining and Zhou, Changjian and Chan, Mansun},
  journal={J. Mater. Chem. C},
  volume={7},
  number={21},
  pages={6273--6278},
  year={2019},
  publisher={Royal Society of Chemistry}
}

@article{comtet2019wide,
  title={Wide-field spectral super-resolution mapping of optically active defects in hexagonal boron nitride},
  author={Comtet, Jean and Glushkov, Evgenii and Navikas, Vytautas and Feng, Jiandong and Babenko, Vitaliy and Hofmann, Stephan and Watanabe, Kenji and Taniguchi, Takashi and Radenovic, Aleksandra},
  journal={Nano Lett.},
  volume={19},
  number={4},
  pages={2516--2523},
  year={2019},
  publisher={ACS Publications}
}

@article{greten2024strong,
  title={Strong coupling of two-dimensional excitons and plasmonic photonic crystals: microscopic theory reveals triplet spectra},
  author={Greten, Lara and Salzwedel, Robert and G{\"o}de, Tobias and Greten, David and Reich, Stephanie and Hughes, Stephen and Selig, Malte and Knorr, Andreas},
  journal={ACS Photon.},
  volume={11},
  number={4},
  pages={1396--1411},
  year={2024},
  publisher={ACS Publications}
}

@article{hopfield1958theory,
  title={Theory of the contribution of excitons to the complex dielectric constant of crystals},
  author={Hopfield, JJ},
  journal={Phys. Rev.},
  number={5},
  pages={1555},
  year={1958},
  publisher={APS}
}

@article{dorey2025transferable,
  title={Transferable Plasmonic Arrays Enabling Strong Coupling with Layered Perovskites in an Active Diode Architecture},
  author={Dorey, Fabien and Ziegler, Jonas D and Moilanen, Antti J and Hordiichuk, Oleh and Nagamine, Gabriel and Taniguchi, Takashi and Watanabe, Kenji and Norris, David J and Kovalenko, Maksym V and Raino, Gabriele and Novotny, Lukas},
  journal={Nano Lett.},
  year={2025},
  volume={26},
  pages={1616},
  publisher={ACS Publications}
}

@article{moilanen2025electrical,
  title={Electrical control of photoluminescence in {2D} semiconductors coupled to plasmonic lattices},
  author={Moilanen, Antti J and Cavigelli, Moritz and Taniguchi, Takashi and Watanabe, Kenji and Novotny, Lukas},
  journal={ACS Nano},
  volume={19},
  number={4},
  pages={4731--4738},
  year={2025},
  publisher={ACS Publications}
}

@article{kim2016effects,
  title={Effects of plasma treatment on surface properties of ultrathin layered {MoS}$_2$},
  author={Kim, Suhhyun and Choi, Min Sup and Qu, Deshun and Ra, Chang Ho and Liu, Xiaochi and Kim, Minwoo and Song, Young Jae and Yoo, Won Jong},
  journal={2D Mater.},
  volume={3},
  number={3},
  pages={035002},
  year={2016},
  publisher={IOP Publishing}
}

@article{wu2019high,
  title={High-performance p-type {MoS}$_2$ field-effect transistor by toroidal-magnetic-field controlled oxygen plasma doping},
  author={Wu, Shaoxiong and Zeng, Yang and Zeng, Xiangbin and Wang, Shibo and Hu, Yishuo and Wang, Wenzhao and Yin, Sheng and Zhou, Guangtong and Jin, Wen and Ren, Tingting and others},
  journal={2D Mater.},
  volume={6},
  number={2},
  pages={025007},
  year={2019},
  publisher={IOP Publishing}
}

\end{document}